\documentclass[aps,prb,twocolumn,superscriptaddress,showpacs,floatfix]{revtex4-2}
\usepackage{amssymb}    
\usepackage{amsmath} 
\usepackage{graphicx}   
\usepackage{verbatim}   
\usepackage{color}      
\usepackage{hyperref}   
\usepackage{epsfig}
\usepackage[utf8]{inputenc}
\usepackage{array}
\usepackage{makecell}
\usepackage{multirow}
\usepackage{gensymb}
\usepackage{float} 
\usepackage[bottom]{footmisc}
\usepackage{braket}
\usepackage{mathtools}
\usepackage{soul}
\usepackage[verbose]{placeins} 
\usepackage{cleveref}
\usepackage{bm}
\usepackage{lipsum}
\usepackage{booktabs} 
\usepackage{csquotes}
\usepackage{booktabs}
\begin{document}


\title{Data-driven high-throughput search for the accelerated discovery of rare-earth-free permanent magnets}

\author{Junaid Jami} 
\email{junaid.jami777@gmail.com}
\affiliation{AbCMS Lab, Department of Metallurgical Engineering and Materials Science, Indian Institute of Technology Bombay, Maharashtra, 400076, India}
\author{Nitish Bhagat} 
\affiliation{AbCMS Lab, Department of Metallurgical Engineering and Materials Science, Indian Institute of Technology Bombay, Maharashtra, 400076, India}
\author{Amrita Bhattacharya} 
\email{b\_amrita@iitb.ac.in} 
\affiliation{AbCMS Lab, Department of Metallurgical Engineering and Materials Science, Indian Institute of Technology Bombay, Maharashtra, 400076, India}

\date{\today}

\begin{abstract}

An integrated data-driven approach combined with a high-throughput framework based on first-principles calculations was used to discover novel rare-earth-free permanent magnets, focusing on binary alloys. Compounds were screened systematically based on their elemental composition, structure, stability, and magnetization. Density functional theory (DFT) calculations were performed on the selected candidates to evaluate their magnetocrystalline anisotropy energy (MAE) and Curie temperature (T$_\mathrm{C}$), resulting
in the identification of ten promising materials. A thorough literature review was done to assess reports of prior existence, which confirmed the novelty of ZnFe and Fe$_8$N. Their ferromagnetic ground state was re-established through DFT, and structural stability was confirmed via negative formation enthalpies, phonon spectra, and elastic criteria. Tetragonal ZnFe and Fe$_8$N exhibit high saturation magnetization ($>$1 T), large anisotropy constants ($>$0.5 MJ/m$^3$), and high T$_\mathrm{C}$ ($>$1200 K). Their magnetic hardness parameters ($\kappa$ = 0.85 for ZnFe and 0.70 for Fe$_8$N) further support their potential as gap magnets. These findings highlight the efficacy of our high-throughput screening, which may serve as a theoretical blueprint for the experimental realization of these materials.

\end{abstract}

\maketitle

\section{Introduction}

The major eras of human civilization are defined by the dominant materials of their time$-$stone, bronze, and iron$-$each marking a disruptive shift in societal development \cite{young2018data}. The desire for new and more efficient materials is driven by the energy requirements of a growing world, necessitating superior energy harvesting and storage solutions \cite{Zhang_2021}. The advancement of modern technologies, from household appliances to cutting-edge microelectronics and eco-friendly innovations like hybrid vehicles and wind turbines, relies on the discovery and optimization of high-performance magnetic materials to meet the growing demands for sustainable energy solutions \cite{coey2011advances, gutfleisch2011magnetic, mccallum2014practical}. However, the downside is that the  discovery and optimization of these materials is an inherently complex challenge that often is slow and labor-intensive, hindered by scattered literature, limited guiding theories, and the intricacies of synthesis, where each candidate material is associated with an almost infinite parameter space of variables that require careful exploration and refinement \cite{maqsood2024future, mal2024leveraging}. Additionally, considerations such as environmental impact, circularity, and sustainability are becoming critical objectives in designing functional material systems. The rarity of elements capable of sustaining magnetic order adds to the complexity and allure of magnetism, where serendipity often drives breakthroughs in discovering high-performance magnets. Notably, the global market for soft and permanent magnets (PMs), valued at \$32.2 billion in 2016, was projected to grow to \$51.7 billion by 2022 \cite{Zhang_2021}. While progress has historically relied on human ingenuity, curiosity and experimentation, it now benefits from advanced tools, computing power, and growing knowledge.

Considering key factors such as elements essential for ferromagnetism, raw material economy, and ecological impact, a novel PM material could involve combinations of four transition metals (Fe, Co, Ni, Mn), seven rare-earth (RE) elements (La, Ce, Y, Pr, Nd, Sm, Dy), and 41 additives. This combinatorial space includes nearly a million binary to quinary RE-free and RE-containing systems with potential for novel PM phases \cite{goll2015novel}. However, traditional investigative approaches, which often demand 2–3 years of focused effort per system, render the identification of even a single promising candidate comparable to searching for a needle in a haystack. To expedite screening, a more advanced and efficient approach becomes imperative.  While first-principles methods are invaluable for predicting material properties with high accuracy, they often struggle when addressing the vast and intricate landscape of material exploration. To complement these methods, semi-empirical strategies like data mining and machine learning have emerged as powerful techniques \cite{huber2020machine,maqsood2024future}, leveraging known computational and experimental data to establish rules, predict properties of unknown materials, and construct quantitative structure-property correlation \cite{liu2019selecting,kusne2014fly}.  Recent advancements in data mining have not only improved criteria for new material formation but also streamlined the prediction of optimal material properties with improved efficiency  \cite{lu2017data}. In computational materials science, data mining has achieved notable success, including the suggestion of new materials in reduced dimension, particularly two-dimensional ones \cite{lebegue2013two, eriksson2018searching}, predicting crystal structures \cite{fischer2006predicting}, discovering new materials \cite{hautier2011data}, and investigating single molecule magnets \cite{dam2014data}. Additionally, it has been instrumental in selecting suitable doping elements for advanced magnetic materials \cite{liu2019selecting} and enhancing predictions of molecular atomization energies by combining data mining techniques with quantum mechanical calculations \cite{rupp2012fast,hansen2013assessment}. 
 
The efficient discovery of novel, commercially viable hard magnetic compounds necessitates well-defined criteria for selecting element combinations. Current widely used PMs, such as Nd$_2$Fe$_{14}$B and SmCo$_5$, rely on RE elements \cite{goll2014high}. The heavier RE elements critical for superior magnetic properties (e.g., Pr, Nd, Sm, Tb, Dy) \cite{skomski1999permanent} are mined using methods with significant ecological and economic challenges, compounded by a fragile global supply chain \cite{gutfleisch2011magnetic}. Hard ferrites, free of RE elements, offer an alternative with high uniaxial magnetocrystalline anisotropy due to their low crystal symmetry, but their performance, up to 20 times weaker than Nd–Fe–B, remains constrained by their lower Fe content and ferrimagnetic behavior. This has spurred extensive research into developing high-performance PMs with reduced or no RE content, including \enquote{gap magnets} that bridge the performance-cost gap between hard ferrites and RE-based PMs \cite{stegen2015heavy,coey2020perspective,skomski2013predicting,lee2020anisotropic}.   
This study emphasizes the selection of elements based on their magnetic properties, economic feasibility, resource availability, and environmental sustainability. Given the vast number of unexplored alloy systems, there remains a strong likelihood of discovering novel materials with promising characteristics. Key screening criteria for potential PMs include intrinsic magnetic properties of their crystal structures, such as saturation magnetization ($M_s$), Curie temperature (T$_{\mathrm{C}}$), and magnetocrystalline anisotropy energy (MAE), which directly influence performance.

 \begin{figure}
    \centering
    \includegraphics[width=1\linewidth]{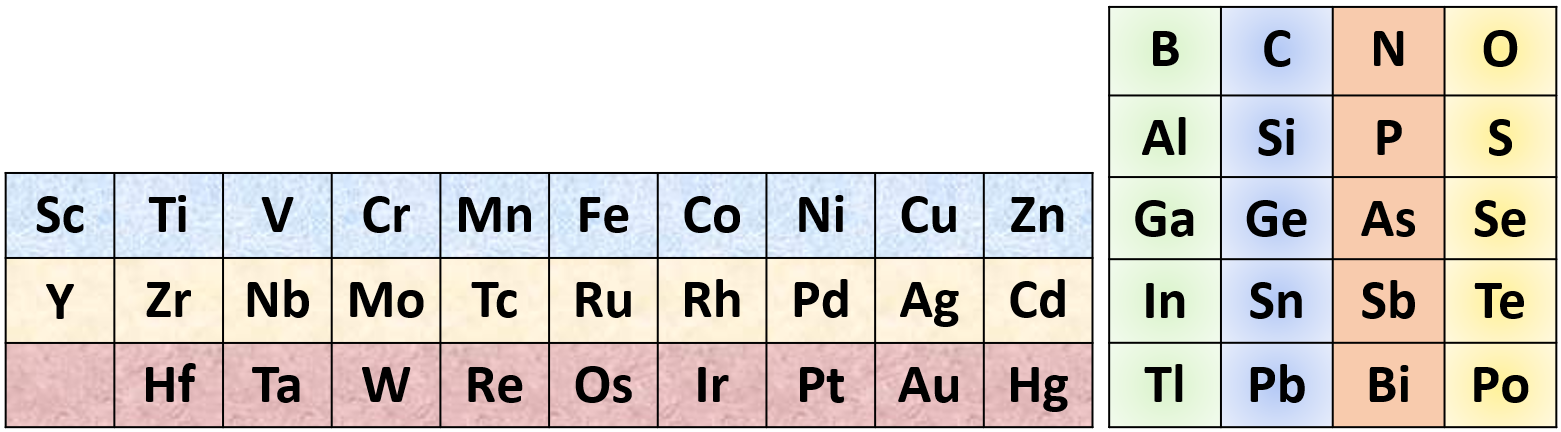}
    \caption{The elements in the compositional space considered in this study. Rare-earth-free binary compounds with a 3$d$ element paired with another 3$d$, 4$d$, 5$d$, or $p$-block element (excluding halogens and noble gases) were filtered from existing database, resulting in 8,372 binary compounds.}
    \label{fig:1}
\end{figure}

Computational materials design, leveraging \emph{ab initio} methods with high-throughput data mining \cite{morgan2004high}, has become a transformative approach for discovering novel and previously unknown phases \cite{MAL2024171590, vishina2020high, vishina2021data}.  Current high-throughput density functional theory (DFT) studies on RE-free permanent magnetic materials focus on tailored screening criteria to identify promising candidates \cite{zhou2023high, nieves2019database, kovacs2020computational, kusne2014fly}. For instance, Vishina \emph{et al.} in \cite{vishina2023fe2c} focused on materials composed of one 3$d$ transition metal (Cr, Mn, Fe, Co, Ni) paired with a $p$-block element (excluding oxygen and noble gases) up to Bi. After applying rigorous filters, this effort yielded three candidates—Fe$_2$C, Mn$_2$MoB$_4$, and Mn$_2$WB$_4$—that hold promise for bridging the performance gap between RE-based and RE-free PMs. In a complementary effort, Mal \emph {et al.} \cite{mal2024leveraging} employed machine learning models to efficiently navigate material spaces, identifying candidates with properties tailored for PM applications. This effort screened 686 materials, ultimately identifying 21 candidates that satisfied criteria for stability, high saturation magnetization, and significant anisotropy constants. However, one of the critical parameters, viz. the Curie temperature,  remains unaddressed in their study. Building on earlier work involving 3$d$–5$d$ compounds \cite{vishina2020high}, which identified compounds based on expensive elements like platinum, Vishina and co-authors later pivoted to RE-free magnets composed of two distinct 3$d$ elements, capitalizing on their typically high saturation magnetization and elevated Curie temperatures, predicting Co$_3$Mn$_2$Ge as a strong candidate due to its high magnetization and elevated T$_{\mathrm{C}}$ \cite{vishina2021data}.  In a really interesting work by Zhou \emph{et al.} \cite{zhou2023high}, they have conducted an efficient search for PM materials using the Inorganic Crystal Structure Database (ICSD), focusing on compounds containing 3$d$ transition elements at specific Wyckoff positions. These positions were selected for their ability to enhance spin-orbit coupling (SOC) through partially occupied orbital multiplets. This systematic approach led to the identification of five promising PM candidates- Fe$_2$Ge, Fe$_2$Sc, Mn$_5$PB$_2$, FeB$_4$, and FeB$_2$. Kusne \emph {et al.} \cite{kusne2014fly} performed high-throughput experimental search of novel RE-free PMs in the Fe-Co-X (X is the transition-metal element) alloys system and suggested Fe$_{78}$Co$_{11}$Mo$_{11}$ to be a strong candidate with the estimated perpendicular anisotropy around 3.6 $\times$ 10$^6$ erg/cm$^3$ (27.0 meV/atom) which is of the same order as that of the Co-Pt alloy \cite{weller1992magnetic}. In another extensive study, Vishina \emph{et al.} \cite{vishina2023stable} mined over one million entries to uncover four promising compounds—Ta$_3$ZnFe$_8$, AlFe$_2$, Co$_3$Ni$_2$, and Fe$_3$Ge—meeting criteria for magnetic performance and metastability. Collectively, these efforts underscore the effectiveness of integrating high-throughput screening, data-driven models, and \emph{ab initio} calculations, which has transformed the discovery of novel RE-free PMs by enabling rapid exploration of vast chemical and structural spaces.

In this work, we employ a combination of DFT, high-throughput searches using the Materials Project database \cite{jain2013commentary} (as of March 2024), and systematic data filtering approach to identify novel RE-free binary alloys as potential PMs.  Transition-metal-based materials, particularly those incorporating 3$d$ elements, are known to provide high saturation magnetization and large Curie temperature—key requirements for PMs. Accordingly, we focused on combinations of 3$d$ elements with other 3$d$, 4$d$, 5$d$, and $p$-block elements (excluding halogens and noble gases) to search for stable binary alloys exhibiting hexagonal or tetragonal crystal structures. Our investigations identified tetragonal ZnFe and Fe$_8$N as promising candidates for RE-free PMs. ZnFe exhibits a high magnetization of 1.15 T, uniaxial anisotropy of 0.75 MJ/m$^3$, a Curie temperature of 1230 K, and a magnetic hardness parameter of 0.84. Similarly, Fe$_8$N demonstrates a saturation magnetization of 1.21 T, uniaxial anisotropy of 0.57 MJ/m$^3$, and a Curie temperature of 1585 K. Data-driven methodologies enable the systematic refinement of candidate materials, effectively reducing the reliance on conventional experimental and computational techniques. By narrowing the search space, these approaches significantly decrease both the cost and time required for the development of novel materials.

\section{Computational Details}
\label{metho}

\begin{figure}
    \centering
    \includegraphics[width=1\linewidth]{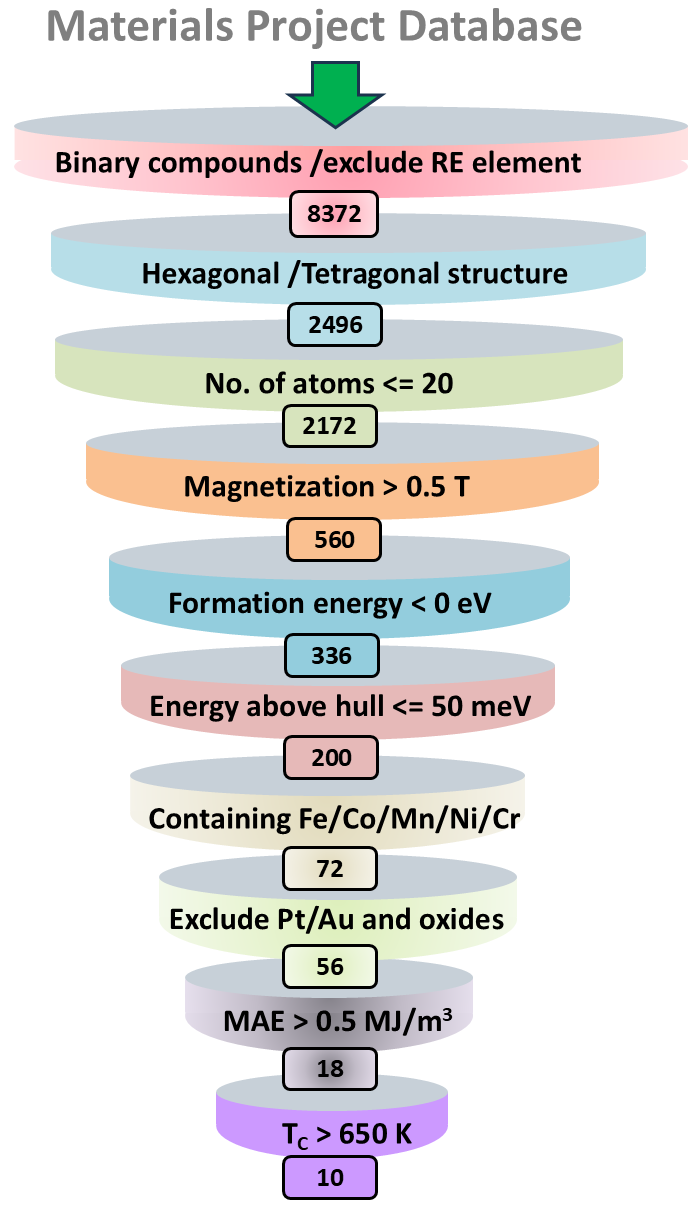}
    \caption{Workflow showing the steps of filters used for our high-throughput search leading to the screening of potential rare-earth-free binary alloys for permanent magnet applications. At each step, the number of potential compounds are filtered based on the applied criteria.}
    \label{fig:2}
\end{figure}

A dataset comprising approximately 20{,}000 calculations on binary compounds from the Materials Project database \cite{jain2013commentary} served as the foundation for our preliminary screening of RE-free candidates. The overall screening workflow, including the sequential application of various filters, is illustrated in Fig.~\ref{fig:2}. Materials Project provides DFT-computed properties such as relaxed structures, atom count per unit cell, magnetization per unit volume, formation energy, and energy above the convex hull. For the initial high-throughput screening, all compounds were assumed to exhibit a ferromagnetic ground state, and the corresponding magnetization values---available from the Materials Project database---were utilized as a key filtering parameter to narrow down potential candidates. Following a multi-stage filtering process, as discussed in detail in section \ref{Screen}, we identified 56 RE-free and earth-abundant compounds crystallizing in hexagonal or tetragonal structures. To advance the screening process, we perform computationally intensive first-principles calculations on the shortlisted compounds, which serve as the basis for applying the next set of filters and conducting further analysis. These calculations encompass the evaluation of MAE and T$_{\mathrm{C}}$, which are critical magnetic performance metrics. The detailed methodologies for the MAE and T$_{\mathrm{C}}$ calculations are discussed in the subsequent paragraphs of this section. Subsequently, for compounds exhibiting the desired set of properties, additional evaluations are conducted to determine the true magnetic ground state, dynamic stability, and mechanical stability, among other relevant characteristics. For this purpose, DFT \cite{PhysRev.136.B864, PhysRev.140.A1133} calculations were carried out within the framework of the Vienna \emph{ab initio} Simulation Package (VASP) \cite{PhysRevB.54.11169}. The exchange and correlation effects between the electrons were treated by employing the projector-augmented wave (PAW) potentials \cite{PhysRevB.50.17953, PhysRevB.59.1758} using the parametrization given by Perdew, Burke, and Ernzerhof within the generalized gradient approximation (GGA) \cite{PhysRevLett.77.3865}. For visualization of crystal structure, the VESTA code \cite{momma2011vesta} is used. 
 
Despite significant advances in local-spin-density electronic structure theory and computational power over the past decades, the accurate determination of MAE still remains difficult and computationally demanding. Traditionally, MAE is evaluated by comparing the total energies of a system for two distinct magnetization orientations—typically in-plane and out-of-plane directions. However, due to the weak spin-orbit coupling (SOC) in 3$d$ transition metals, the force theorem \cite{daalderop1990first} is often employed, wherein MAE is estimated by comparing the band energies for two different magnetization orientations. The main difficulty associated with this method concerns the numerical stability of calculating a very small difference of two large numbers.  Consequently, eliminating numerical fluctuations requires exceptionally fine k-point sampling for accurate integration over the Brillouin zone \cite{wang1996torque}. Experimentally, the MAE can be determined using torque magnetometry \cite{o2000modern} and a similar approach can be employed in \emph{ab initio} calculations, where MAE is obtained through magnetic torque computations \cite{staunton2006temperature, bornemann2007magnetic}. This method offers clear advantages over total energy calculations, as it allows for the direct determination of MAE from a single calculation, eliminating the need to accurately extract a small energy difference between two large values and thereby circumventing numerical challenges.  In this work, MAE was calculated using the torque method as implemented in the spin-polarized relativistic Korringa-Kohn-Rostoker (SPRKKR) package \cite{ebert2011calculating}, following the formalism of \cite{staunton2006temperature}. The technologically relevant quantity, that is the MAE constant ($K$), can then be calculated as the MAE per unit volume $(V)$. 
\begin{equation}
	\label{eqn:E2}
	K=\frac{\mathrm{MAE}}{V}
\end{equation}
A positive value of MAE or $K$ corresponds to an uniaxial anisotropy which means the easy magnetization axis is perpendicular to the $ab$ plane i.e., along the $c$-axis, and therefore indicates the material to be suitable for PM application.

\begin{figure*}[t]
    \centering
    \includegraphics[width=1\linewidth]{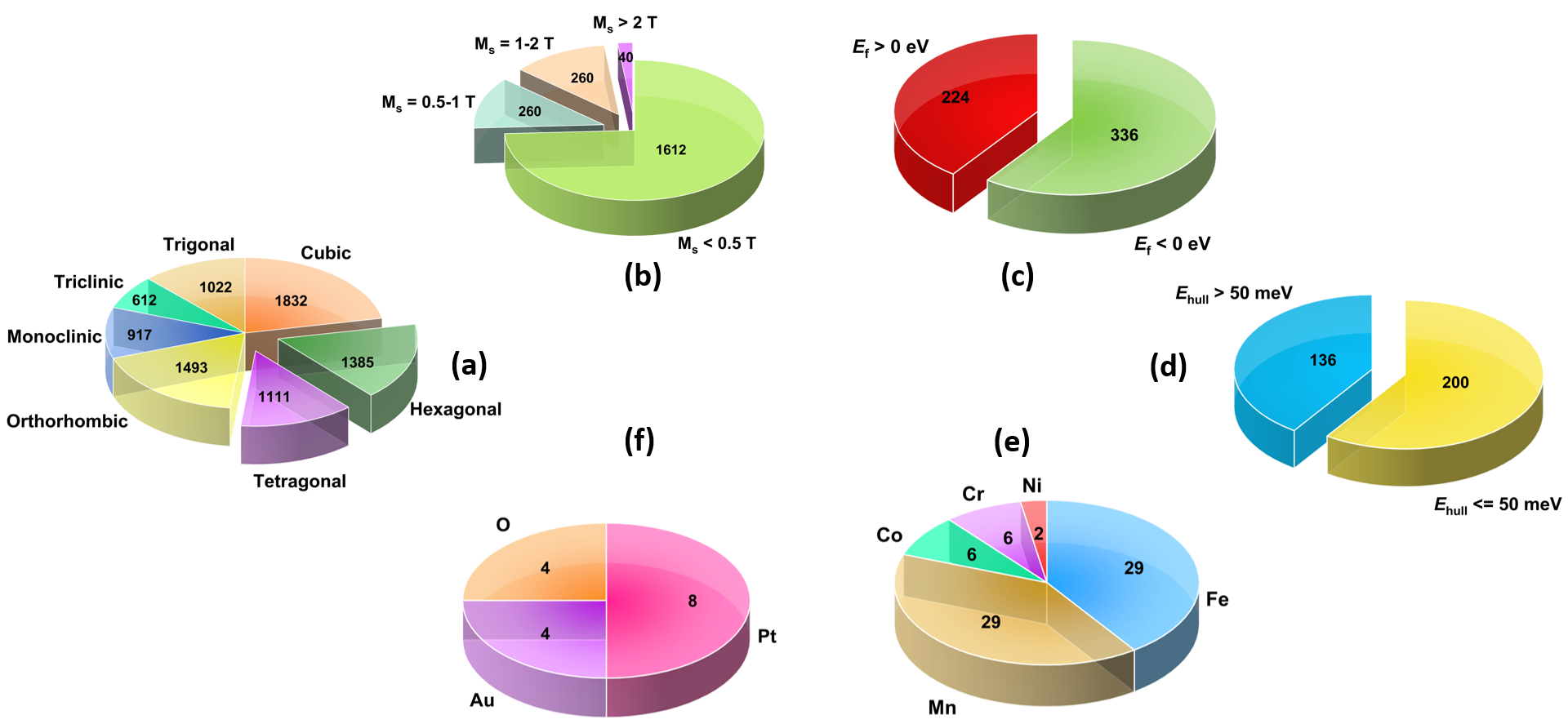}
    \caption{ Pie charts (a)–(f) illustrating the output of our high-throughput screening at different stages. (a) Distribution of the 8372 compounds according to their structures. Out of the seven different crystal structures, only hexagonal (1385) and tetragonal (1111) structures were considered for proceeding to the next step; which may have non zero structural anisotropy. (b)  Subsequently, a magnetization filter of $M_s > 0.5$ T  was applied for the structures with $\leq$ 20 atoms in the unit cell, following which 1612 compounds were removed. A total of 260 compounds exhibited magnetization between 0.5–1 T, another 260 exhibited within 1–2 T, and only 40 surpassed 2 T. (c) Out of these compounds, the stable ones were screened based on negative formation energy and 336 compounds were found to be stable, while 224 were deemed unstable. (d) Following this, an additional stability criterion, energy above hull ($E_{\text{hull}} \leq $ 50 meV)  was applied, which further refined the dataset to 200 compounds. (e) A subset of 72 compounds containing at least one of Fe, Co, Ni, Mn, or Cr was selected. (f) Finally, the oxide compounds and the compounds containing costly elements such as Pt and Au were excluded, yielding 56 candidates for proceeding further.}
    \label{fig:3}
\end{figure*}

The Heisenberg exchange coupling parameters $J_{ij}$ were calculated by mapping the system onto a Heisenberg Hamiltonian, using the method proposed by Liechtenstein \emph {et al.} \cite{liechtenstein1987local}, as implemented in the SPRKKR package. The calculations were carried out within the scalar relativistic mode and for consistency, the same functional (GGA-PBE) for exchange and correlation has been chosen as before \cite{PhysRevLett.77.3865}.  The angular momentum expansion up to $l_{max}$ = 3 has been taken for each atom to ensure the accurate description of the electronic structure near the atomic sites and the integration over the Green’s functions was carried out with 30 energy points. In order to further improve the charge convergence with respect to $l_{max}$,  Lloyd’s formula has been employed for determining the Fermi energy \cite{doi:10.1080/00018737200101268, Zeller_2008}. The convergence criterion was set at level of $10^{-5}$ and 10,000 k-points were employed per reduced Brillouin zone. The exchange coupling parameters are calculated with respect to the central site $i$ for cluster atoms with radius $R_{clu} = \mathrm{max}|R_i-R_j|$. We have taken the radius of a sphere $R_{clu}$ of 7.0. The Curie temperature, is then calculated within the mean-field approximation using the approach discussed in \cite{PhysRevB.108.054431}.

In high-throughput studies, the initial candidate pool is typically refined by evaluating the thermodynamic stability of compounds, commonly assessed through the formation energy and the energy above the convex hull, which quantify a material’s tendency to decompose into its constituent elements or competing phases. After the final selection of promising compounds, dynamic stability was evaluated via phonon calculations within the harmonic approximation. Force constants were computed using density functional perturbation theory (DFPT) within VASP, and phonon spectra were subsequently obtained using the PHONOPY package \cite{TOGO20151}. In addition to dynamic stability, elastic and mechanical properties were analyzed to assess the structural robustness and technological viability of the selected materials. Mechanical stability is governed by the elastic constants $C_{ij}$, which relate stress to strain in a homogeneous medium. The full elastic coefficient matrix was calculated using VASP, and the mechanical stability of each crystal class was evaluated based on the Born–Huang criteria \cite{born1996dynamical}.

An estimation of the potential of these materials for PMs can be made by evaluating key intrinsic hard magnetic properties. Besides having a high T$_{\mathrm{C}}$, a PM material must also exhibit a high energy product  $(BH)_{max}$, the maximum of the product of the magnetic field induction $B$ and the magnetic field $H$ in the second quadrant of the $BH$ curve.

\begin{equation}
\label{eqn:E3}
   (BH)_{max} = \mu_0 {M_s{^2}}/4
\end{equation}

$(BH)_{max}$ represents the theoretical upper limit assuming an ideal rectangular $M-H$ loop. Here, $\mu_0$ is the vacuum permeability, $M_s$ is the saturation magnetization of the material. Even though a large magnetization is desirable for PMs, the ratio between magnetization and MAE is even more important since it determines the magnetic hardness parameter ($\kappa$) \cite{1130282271085861760,6008648}, a dimensionless quantity used to classify ferromagnetic materials  as hard, semi-hard, or soft. 

\begin{equation}
\label{eqn:E4}
  \kappa   =\sqrt{\frac{K}{\mu_0 M_s^{2}}} 
\end{equation}

A value of $\kappa\geq1$ represents hard magnets, and $\kappa \geq 0.1$ indicate semi-hard magnets. PMs fall within the hard or semi-hard range with $\kappa > 0.1$.

\section{High-throughput and Screening}
\label{Screen}

For a material to qualify as a strong PM, it must exhibit ferromagnetic ordering with a high saturation magnetization ($M_s$ $>$ 1 T), a large Curie temperature (T$_{\mathrm{C}}$  $>$ 550 K), and most critically, a large uniaxial MAE ($K$ $>$ 1 MJ/m$^3$). The search conducted in this study followed a two-stage approach to identify suitable candidates. In the first stage, as depicted in Fig. \ref{fig:3}, a high-throughput screening was performed on a vast number of potential compounds to reduce the dataset before moving to more computationally demanding calculations. This step utilized several screening criteria that relied solely on data from the Materials Project database \cite{jain2013commentary} , without involving any theoretical calculations. In the second stage, the selected systems underwent high-precision, computationally expensive calculations to determine their magnetic anisotropy and ordering temperature.

As illustrated in Fig.~\ref{fig:1}, the initial stage of material selection focused on binary compounds comprising at least one 3$d$ transition metal, combined with either another 3$d$, 4$d$, 5$d$, or $p$-block element (excluding halogens and noble gases). This choice was motivated by the fact that 3$d$ elements typically exhibit higher saturation magnetization and larger Curie temperatures compared to their 4$d$ and 5$d$ counterparts \cite{khomskii2014transition}. To prioritize the identification of RE-free PMs, all compounds containing RE elements were excluded due to the economic and environmental concerns associated with RE mining and processing. The overall workflow of the high-throughput screening strategy is presented in Fig.~\ref{fig:2}. Following the first filtering stage which involved elemental selection criteria, the dataset was reduced to 8,372 compounds, selected from an initial pool of over 20,000 binary compounds. In the next step, all compounds with cubic symmetry were excluded, as spin-orbit coupling effects are known to influence the magnetic anisotropy energy most effectively in non-cubic compounds \cite{PhysRevB.39.865}. Therefore, avoiding cubic structures should be a guiding principle in the search for strongly anisotropic magnetic materials. We specifically screened for tetragonal and hexagonal systems, as their low symmetry often leads to higher magnetocrystalline anisotropy \cite{fazekas1999lecture}, and restricted our analysis to these candidates. It is worth noting that the majority of well-established permanent magnetic materials belong to these two crystal systems. For instance, hexagonal structures are found in Co$_3$Pt, MnBi, BaFe$_{12}$O$_{19}$, and SmCo$_5$, whereas tetragonal symmetry is exhibited by FePt, CoPt, MnAl, and Nd$_2$Fe$_{14}$B \cite{o2000modern}.  We applied an additional filter to exclude materials with more than 20 atoms per unit cell, as larger unit cells significantly increase the computational cost of high-precision calculations. For PMs, a high saturation magnetization is essential as it directly influences the magnet's strength and sets the upper limit for remanence. To identify potential candidates, we applied a screening criterion of $M_s>$ 0.5 T, ensuring the selection of materials with significant magnetic strength. 

A key step in data mining and high-throughput searches for materials design is evaluating the stability of compounds to identify suitable candidates for experimental synthesis. In this study, stability is assessed using thermodynamic criteria, such as the formation energy and the compound’s position relative to the convex hull of competing phases. We considered only compounds with negative formation energies and filtered those within 50 meV/atom from the convex hull, significantly enhancing their likelihood for experimental realization. Notably, convex hull evaluation can reduce the number of stable compounds by one order of magnitude \cite{Zhang_2021}. After performing the stability analysis, we narrowed the dataset down to 220 viable compounds. This methodology not only ensures desirable magnetic properties but also enhances the practical feasibility of synthesizing the proposed compounds. 

Compounds containing at least one of the 3$d$ transition metals viz. Fe, Co, Ni, Mn, or Cr - were selected for further analysis. During this process, several oxide materials were identified; however, transition metal oxides, known for their strongly correlated electronic structures, pose challenges for single-determinant theories like DFT in accurately describing their properties \cite{vishina2020high}, leading to their exclusion from our study. It is important to clarify that their exclusion was not due to the lack of their potential as PMs, but rather because these materials exhibit complex and intricate magnetic behaviors that cannot be generalized with a single theoretical framework, especially when it comes to accurately predicting MAE and other magnetic properties. 

The precise calculation of MAE remains one of the most challenging aspects of such studies. In this work, we set the threshold value of $K$ to 0.5 MJ/m$^3$ and compounds having higher $K$ values than this were selected for subsequent analysis. Furthermore, while extrinsic properties like remanent magnetization and coercivity are strongly influenced by the microstructure of the material, their upper limits are fundamentally dictated by intrinsic quantities such as the $M_s$ and $K$. High-performance PMs are typically expected to have T$_{\mathrm{C}}$ values exceeding 550 K, as outlined by Coey \cite{COEY2012524}. In this study, the T$_{\mathrm{C}}$ of the compounds in their ground-state ferromagnetic (FM) configuration was calculated using exchange coupling parameters within the framework of the mean-field approximation. Although the mean-field approximation is computationally efficient, it is well-documented to generally overestimate T$_{\mathrm{C}}$ values \cite{PhysRevB.64.174402}.  To mitigate this limitation and ensure practical applicability, we employed a stricter filtering criterion by setting the T$_{\mathrm{C}}$ cutoff at 650 K, providing a margin of safety in our predictions. 

\begin{figure}[t]
    \centering
    \includegraphics[width=1\linewidth]{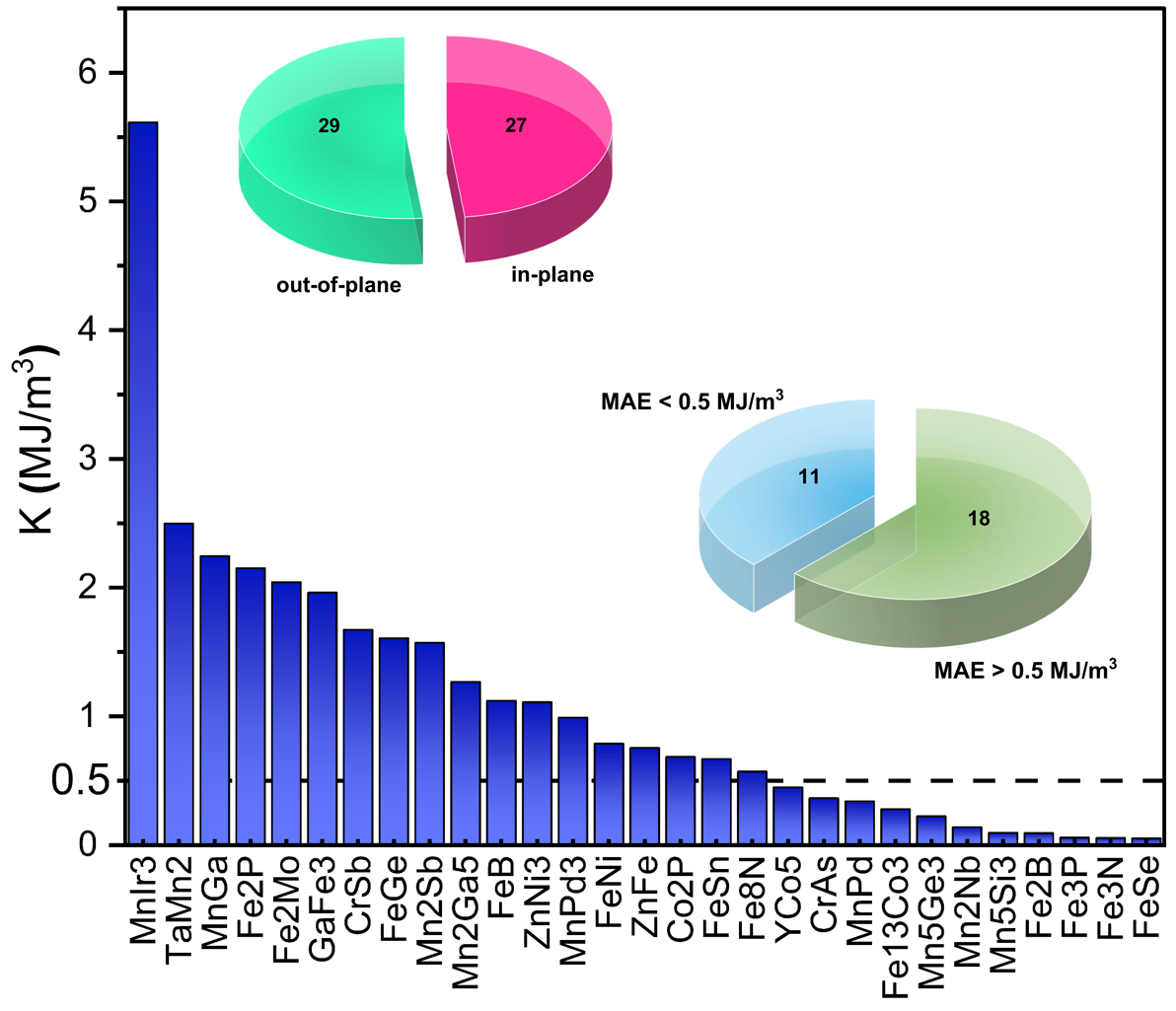}
    \caption{Magnetic anisotropy energy (K) was calculated for all the compounds retained after applying the previous filters. Out of these 56 compounds, 29 compounds exhibited out-of-plane anisotropy and their K is compared using bar chart, while remaining were found to have in plane anisotropy. The horizontal dashed line at 0.5 MJ/m$^3$ in the bar chart represents the screening threshold for potential permanent magnet candidates, leading to the selection of 18 compounds for further computation.}
    \label{fig:4}
\end{figure}

The most widely used RE PMs, Nd$_2$Fe$_{14}$B and Sm$_2$Co$_{17}$, possess magnetic anisotropy energy density ($K$) values of 4.9 MJ/m$^3$ and 3.3 MJ/m$^3$, respectively, along with saturation magnetization $M_s$ of 1.6 T and 1.3 T, and Curie temperatures T$_{\mathrm{C}}$ of 585 K and 1100 K \cite{liu2006overview,coey1996rare,CUI2018118}. In the present study, less stringent initial screening criteria (like $M_s > 0.5$ T and $K>$ 0.5 MJ/m$^3$) than the practical performance requirements for high-performance PMs, were deliberately set for filtering possible PM candidates. This relaxed filtering criterion ensures that potentially promising new classes of compounds are not prematurely excluded because some materials may exhibit significant improvements in magnetic properties after alloying or structural refinements. Following the final stage of screening, a comprehensive literature survey was undertaken to cross-reference the identified compounds and evaluate their prior theoretical or experimental reports. This assessment aimed to determine whether any of the shortlisted compounds had already been proposed as potential PMs, were known to exhibit non-ferromagnetic behavior, or possessed structural classifications that remain ambiguous, thereby indicating potential challenges in their realization in the laboratory. Compounds that were found to be well-known, extensively studied, or associated with any significant experimental or theoretical limitations were excluded from the list. Consequently, only those compounds that appeared to be both potentially viable and previously unexplored were retained for subsequent investigation.

\begin{figure}[t]
    \centering
    \includegraphics[width=1\linewidth]{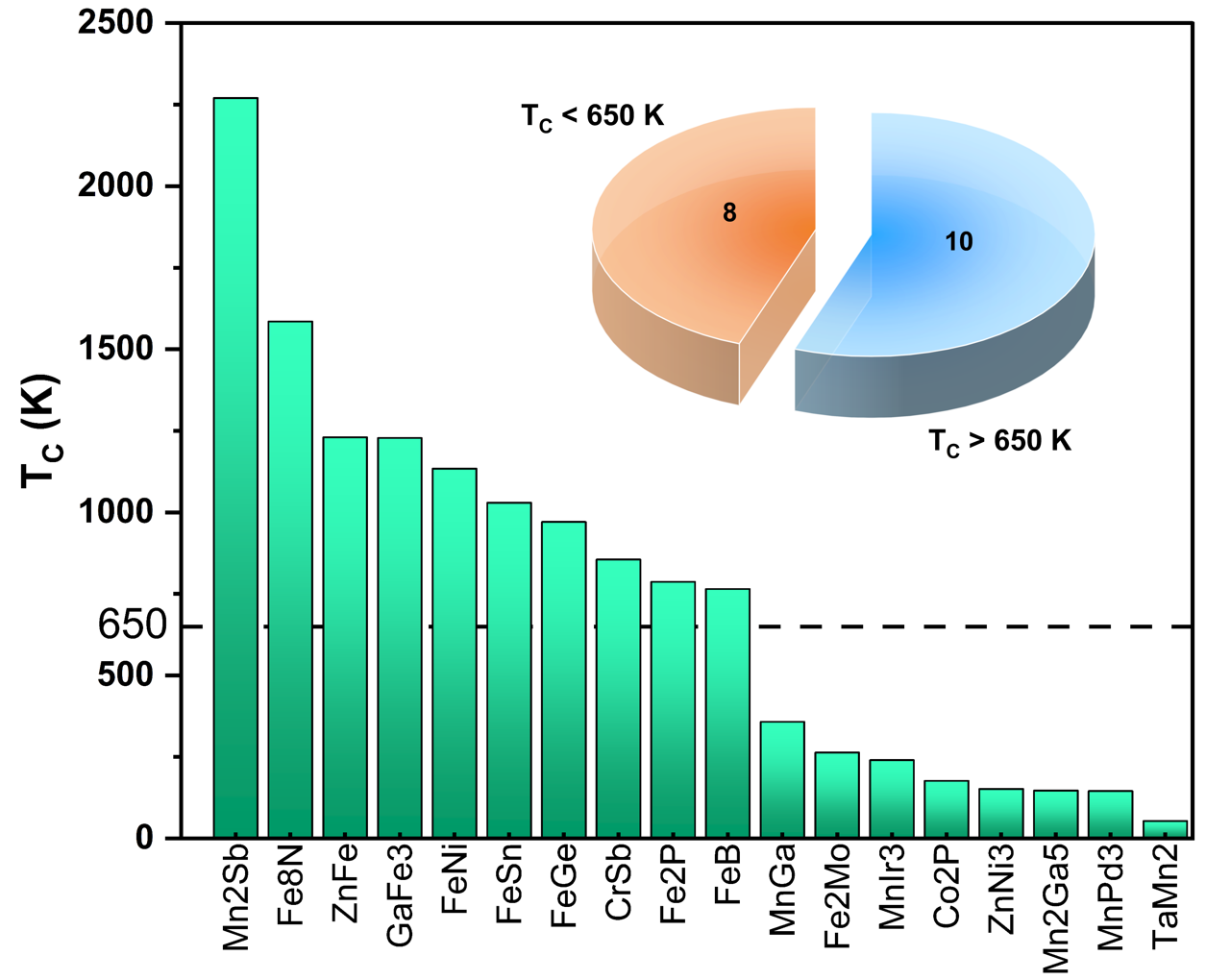}
    \caption{ Bar chart showing the Curie temperature T$_{\mathrm{C}}$ of 18 compounds. A screening threshold of 650 K, which represented the horizontal dashed line, is applied to filter out materials with insufficient thermal stability for high-temperature applications. After this selection criterion, 10 compounds remained as potential candidates for further detailed investigation.}
    \label{fig:5}
\end{figure}

\section{Results and Discussions}

\subsection{Findings from high-throughput calculations}

A few notable inferences can be drawn from the screened dataset. First, compounds containing Fe and Mn generally exhibit larger MAE, with \text{MnIr}$_3$ showing the highest MAE exceeding 5~MJ/m$^3$, followed by \text{TaMn}$_2$, \text{MnGa}, \text{Fe}$_2$\text{P}, and \text{Fe}$_2$\text{Mo}, all of which exhibit MAE values above 2~MJ/m$^3$. Second, Fe-based systems tend to show higher T$_{\mathrm{C}}$, with compounds such as \text{Fe}$_8$\text{N}, \text{ZnFe}, \text{GaFe}$_3$, \text{FeNi}, and \text{FeSn} exhibiting T$_{\mathrm{C}}$ values exceeding 1000~K, indicative of strong magnetic exchange interactions. In contrast, Mn-based compounds generally display lower T$_{\mathrm{C}}$ values, with the notable exception of \text{Mn}$_2$\text{Sb}, which shows the highest T$_{\mathrm{C}}$ in the dataset at 2270~K. Finally, after applying all filtering criteria outlined in Fig.~\ref{fig:2}, 8 out of the 10 final shortlisted compounds suitable for PM applications are Fe-containing, highlighting the key role of Fe in the design of high-performance RE-free magnetic materials. The magnetic parameters of these compounds are summarized in Table~\ref{tab:table1}, some of which have already been reported in previous theoretical and/or experimental investigations. However, a number of additional considerations must be addressed for proposing them as PM candidates. In the Materials Project database \cite{jain2013commentary}, which served as the source of our dataset, all compounds are assumed to possess a FM ground state. It is therefore essential to re-evaluate the stability of the FM state relative to other possible magnetic configurations. Furthermore, the structural stability of each compound must be reassessed in comparison to alternative crystallographic phases. In the following section, we present a comprehensive analysis of each of these compounds based on our literature survey.

\begin{table*}[t!]
    	\caption{ The different parameters for the filtered compounds obtained after the screening steps performed in this investigation viz. their corresponding crystal structure (Tet = Tetragonal, Hex = Hexagonal, Ortho = Orthorhombic), saturation magnetization $M_s$ (T), magnetocrystalline anisotropy constant $K$ (MJ/m$^3$), Curie temperature T$_{\mathrm{C}}$ (K),  hardness parameter $\kappa$,  maximum energy product $(BH)_{max}$ (kJ/m$^3$), and anisotropy field H$_a$ (T).}
	\label{tab:table1}
 
    \begin{tabular}{p{2cm}p{2cm}p{2cm}p{2cm}p{2cm}p{2cm}p{2cm}p{1cm}}
    \hline
	
       Material& Structure& $M_s$ & $K$  & T$_{\mathrm{C}}$ & $\kappa$ & $(BH)_{max}$  & H$_a$\\  
        \hline
         FeSn & Hex & 0.77  & 0.67 & 1029 & 1.19 & 117.71 & 2.18 \\
         CrSb & Hex & 0.81  & 1.67 & 856 & 1.80 & 129.33 & 5.21\\ 
         Mn$_2$Sb & Tet & 1.76 & 1.57 & 2270 & 0.79 & 617.92 & 2.24\\ 
         FeB   & Tet &1.54  & 1.12 & 765 & 0.77  & 469.95 & 1.83 \\
          FeB & Ortho & 1.39  & 0.98 & 552 & 0.40  & 1523.91 & 0.89\\ 
         FeNi & Tet & 1.85 & 0.79  & 1134  & 0.54 & 681.19 & 1.07\\ 
        Fe$_2$P & Hex & 1.08 & 2.15 & 787 & 1.52 & 233.07 & 4.99\\ 
        Fe$_3$Ga & Hex & 1.79 & 1.96 & 1228 & 0.88 & 638.49 & 2.75\\ 
         FeGe & Hex & 0.95 & 1.61 & 971 & 1.50 & 178.29 & 4.27 \\
         ZnFe & Tet & 1.15 & 0.76 & 1230 & 0.85 & 264.13 & 1.65\\ 
         Fe$_8$N & Tet & 1.21 & 0.57 & 1585 & 0.70 & 292.78  & 1.18\\ 
         \hline
    \end{tabular}
\end{table*}

\textbf{FeSn}- Several experimental analysis using the Mössbauer studies \cite{haggstrom1975studies}, a neutron scattering experiment \cite{yamaguchi1967neutron}, and a perturbed angular correlation study \cite{wodniecka2001hyperfine} have been conducted on FeSn. This compound crystallizes in a hexagonal structure and exhibits antiferromagnetic ordering with a Néel temperature of 368 K \cite{giefers2006high, ehret1933x}. Below Néel temperature, FeSn adopts a planar antiferromagnetic configuration, while maintaining ferromagnetic ordering within each kagome type layer \cite{sales2019electronic}.

\textbf{CrSb} - At low temperature, CrSb crystallizes in the hexagonal NiAs-type antiferromagnetic phase  with a Néel temperature of 710 K \cite{radhakrishna1996inelastic, abe1984magnetic}. Theoretically, this NiAs-type structure is found to be the most stable in the antiferromagnetic phase, consistent with experimental observations. \cite{kahal2007magnetic}

\textbf{Mn$_2$Sb} - Mn$_2$Sb has tetragonal Cu$_2$Sb-type crystal structure and neutron diffraction studies on both single-crystal and powdered samples of Mn$_2$Sb have confirmed its ferrimagnetic ordering \cite{wilkinson1957magnetic} with a transition temperature at about 550 K \cite{suzuki1992electronic}. Furthermore, these investigations have experimentally revealed the distinct magnetic contributions from two non-equivalent Mn sites within the compound \cite{KIMURA1992707}.

\textbf{FeB} - FeB is one of the two stable phases in the binary iron-boron system \cite{LI2014211}, crystallizing in an orthorhombic lattice with space group $Pnma$ and eight atoms per unit cell \cite{bjurstrom1929rontgenanalyse}. This phase exhibits a T$_{\mathrm{C}}$ of 598 K and a first-order anisotropy constant of approximately 0.4 MJ/m\(^3\) \cite{zhdanova2013magnetic}. Given its greater stability, we also focused on the orthorhombic structure for our calculations, obtaining a saturation magnetization ($M_s$) of 1.39 T, an anisotropy constant ($K$) of 0.98 MJ/m\(^3\), a T$_{\mathrm{C}}$ of 552 K, and a magnetic hardness parameter ($\kappa$) of 0.4.

\textbf{FeNi} - Tetrataenite (L1$_0$-FeNi) is a promising RE-free PM candidate due to its favorable intrinsic properties. Minute quantities of L1$_0$-FeNi are naturally found in iron meteorites \cite{petersen1977mossbauer}. Despite being composed of abundant elements such as Fe and Ni, it exhibits a high uniaxial magnetic anisotropy, oscillating around 1 MJ/m$^3$ \cite{pauleve1968magnetization}. Its saturation magnetic flux density reaches 1.6 T, comparable to that of Nd-Fe-B, while its T$_{\mathrm{C}}$ exceeds 823 K, surpassing that of conventional magnets \cite{goto2017synthesis}.  FeNi alloys, consisting of two 3$d$ elements, present an attractive alternative, offering a theoretically achievable maximum energy product of 446 kJ/m$^3$ \cite{aladerah2024pressure}. Moreover, mean-field calculations by Kübler \cite{kubler2006ab} predict a T$_{\mathrm{C}}$ of 1130 K, consistent with our findings.  

\textbf{Fe$_2$P} - The magnetic properties of hexagonal single crystals of Fe$_2$P, investigated in \cite{senateur1976etude}, reveal a ferromagnetic ground state with a T$_{\mathrm{C}}$ nearly 210 K. The material exhibits a high magnetic anisotropy constant of $K$ = 2.32 MJ/m$^3$ \cite{fujii1977magnetic, ishida1987electronic} which is in good agreement with our findings.  

\textbf{Fe$_3$Ga} -   Fe$_3$Ga ($\beta$) forms an ordered face-centered cubic structure of the Cu$_3$Au (L1$_2$) type, with an extrapolated T$_{\mathrm{C}}$ of 1040 K, approaching that of pure body-centered cubic iron \cite{kawamiya1982magnetic}. However, other competing phases, such as DO$_3$-type and DO$_{19}$-type Fe$_3$Ga, have also been reported \cite{yan2016temperature, golovin2015structure}.

\textbf{FeGe} - FeGe in the cubic B20 phase is an experimentally well-studied prototypical chiral magnet exhibiting helical spirals and is a candidate material for potential applications in skyrmion based computation, but the T$_{\mathrm{C}}$ is reported to be  close to the room temperature \cite{bak1980theory, xu2017magnetic, grytsiuk2019ab}.

 \textbf{ZnFe} - To the best of our knowledge, the structural and magnetic properties of this material have not been investigated in detail before and are discussed in the subsequent sections.  
 
\textbf{Fe$_8$N ($\alpha''$-Fe$_{16}$N$_2$)} - The magnetic properties of technologically significant ferromagnetic transition-metal compounds can be enhanced by incorporating nitrogen into interstitial sites, as observed in Fe$_8$N. Iron nitride attracts considerable interest because of its exceptionally high magnetization and because of its earth abundant elemental constitution. 
The Fe-N system exhibits a rich variety of binary compounds, each characterized by a distinct nitrogen concentration and magnetic behavior. There exists a series of binary Fe-N compounds:  $\gamma''$-FeN$_y$ ($y = 0.9$--$1.0$) with $\sim$50 atomic \% N, $\zeta$-Fe$_2$N with $\sim$33\% N, $\varepsilon$-Fe$_3$N$_{1+y}$ ($y = 0$--$0.33$) with $\sim$25\% N, $\gamma'$-Fe$_4$N with $\sim$20\% N, and Fe$_8$N (or $\alpha''$-Fe$_{16}$N$_2$) with $\sim$11\% N \cite{tessier2000energetics}. Among these, Fe$_8$N has garnered significant interest due to its combination of high magnetization and considerable magnetocrystalline anisotropy. Since the discovery of the interstitial $\alpha''$-Fe$_{16}$N$_2$ compound, which crystallizes in a body-centered tetragonal structure and exhibits a high saturation magnetization of 2.8 T, it has garnered renewed interest in recent years, particularly after a $K=$1.9 MJ/m$^3$ was reported based on thin-film measurements using the Stoner-Wohlfarth model \cite{jack1951occurrence, MOHAPATRA2023170258}. Fe$_8$N, which adopts a tetragonal structure (space group $I4/mmm$, No. 139) \cite{bao2024fe, timoshevskii2001influence} and the saturation magnetization reported on films, foils and powder are highly scattered with Curie temperature was estimated to be around 813 K by using the Langevin function \cite{sugita1991giant}. However, its bulk magnetic properties remain largely unexplored due to the challenges associated with synthesizing a single-phase bulk material.  Therefore, we have undertaken a detailed investigation of its magnetic characteristics, which are presented in Table \ref{tab:table1} and is discussed in details in subsequent sections.

 With the exception of ZnFe, most of these materials have been extensively studied in the literature and were therefore excluded from further consideration. Beyond the initial discovery, the optimization of such complex materials remains a formidable challenge, predominantly confined to a linear optimization techniques wherein compositional tuning via doping and iterative characterization is employed. This approach, however, is inherently slow and resource-intensive, limiting the rate of material advancement.

\begin{figure*}[t]
    \centering
    \includegraphics[width=1\linewidth]{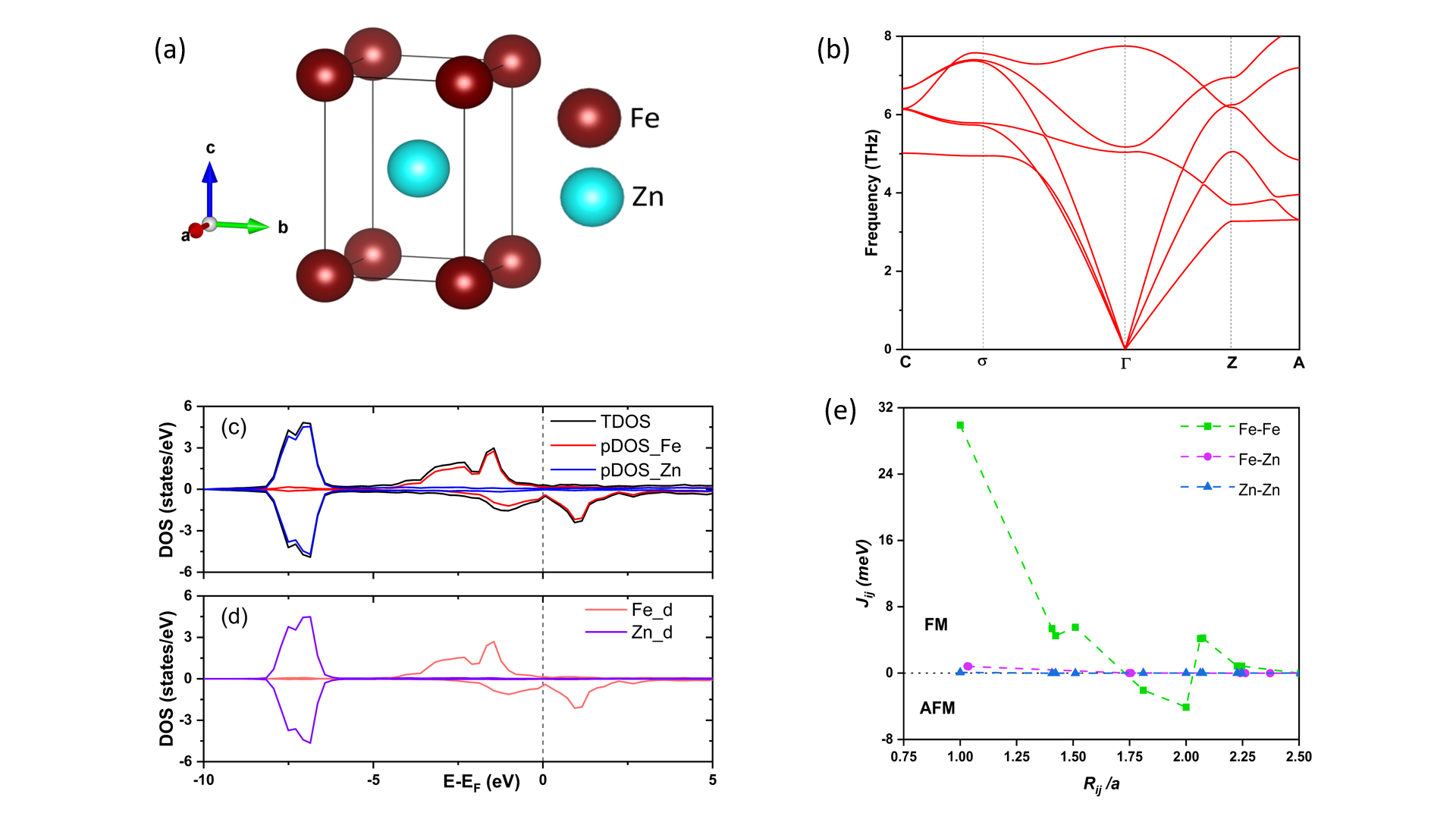}
    \caption{(a) Tetragonal unit cell of ZnFe. (b) Phonon dispersion curves of ZnFe, confirming its dynamical stability due to the absence of imaginary frequencies. (c) Spin-polarized total density of states (TDOS) and atom-projected partial density of states (pDOS) of ZnFe, with the Fermi level ($E_F$) set as the reference. (d) Orbital-resolved contributions of Fe and Zn $d$-orbitals to the TDOS, highlighting their role in the electronic structure. (e) Heisenberg exchange coupling parameters ($J_{ij}$) plotted as a function of the interatomic distance normalized by the lattice parameter ($R_{ij}/a$), providing insights into the magnetic interactions in ZnFe.}
    \label{fig:6}
\end{figure*}

\subsection{Structural stability and magnetic properties of ZnFe}

To the best of our knowledge, there are no prior reports available on the crystal structure or magnetic properties of ZnFe.
Interestingly, our findings reveal that ZnFe possesses all the key magnetic attributes required for potential PM applications, as summarized in Table~\ref{tab:table1}. The accurate determination of the magnetic ground state is particularly crucial in governing the stability of a magnetic compound, as magneto-structural coupling can significantly alter the energy landscape. However, depending on the crystal structure and the nature of the magnetic ions, the number of possible antiferromagnetic configurations that need to be considered can be substantial. As a result, identifying the true magnetic ground state is a crucial challenge in predictive high-throughput screening for magnetic materials. Since many database entries do not explicitly account for the correct magnetic ground state, we conducted an extensive examination of multiple magnetic configurations to accurately determine the ground state of ZnFe.

Given that the ZnFe unit cell contains a single magnetic atom (Fe), a $2\times2\times2$ supercell was constructed to systematically explore the possible ferromagnetic (FM), antiferromagnetic (AFM), and ferrimagnetic (FiM) states. To establish the most stable magnetic ordering, we considered multiple AFM and FiM arrangements by systematically varying the relative spin alignments of Fe atoms within the supercell. Details of the non-degenerate AFM and FiM configurations examined are provided in Section S1 of the Supplementary Information. After performing total energy calculations for all considered magnetic states, the FM configuration was found to have the lowest total energy, confirming it as the magnetic ground state of ZnFe. This result highlights the intrinsic preference of ZnFe for a parallel spin alignment, which is crucial for its potential application in PM development.

The stability of ZnFe was assessed through its formation enthalpy, which was found to be negative ($-0.028$ eV/f.u.) for the FM ground state, thereby confirming the compound's stability. However, while a negative $E_\mathrm{f}$ suggests that the compound is energetically favorable relative to its constituent elements, it does not guarantee experimental realization. Therefore, an additional evaluation of dynamical stability was carried out using phonon dispersion calculations based on DFPT. The resulting phonon spectra, computed along high-symmetry paths in the Brillouin zone and shown in Fig.~\ref{fig:6}(b), reveal no imaginary frequencies. The absence of negative phonon modes confirms that ZnFe is dynamically stable and resistant to local atomic perturbations, reinforcing the feasibility of experimental synthesis.
  
\begin{table*}[t!]
 \caption{Elastic constants ($C_{ij}$) in GPa for ZnFe and Fe$_8$N.}
   \centering

    \setlength{\tabcolsep}{10pt}
    \begin{tabular}{lccccccc}
        \toprule
        Compound & $C_{11}$ & $C_{12}$ & $C_{13}$ & $C_{16}$ & $C_{33}$ & $C_{44}$ & $C_{66}$ \\
        \midrule
        ZnFe  & 173.717 & 122.606 & 122.606 & 0 & 225.551 & 43.090 & 109.691 \\
        Fe$_8$N & 247.278 & 119.479 & 151.891 & 0 & 295.614 & 92.913 & 113.843 \\
        \bottomrule
    \end{tabular}
    \label{tab:elastic_constants}
\end{table*}

Beyond dynamical stability, mechanical stability was also examined, as it describes the structural robustness against elastic deformations. This was evaluated using the Born-Huang stability criteria \cite{born1996dynamical,SINGH2021108068}, which establish necessary conditions for the elastic constants that must be met for a mechanically stable structure. For a tetragonal crystal, these conditions are given by:  $C_{11}-C_{12} > 0$,  $2C_{13}^{2}< C_{33} (C_{11}+C_{12})$, $C_{44} > 0$, and  $2C_{16}^{2}< C_{66} (C_{11}-C_{12})$. The computed elastic constants for ZnFe are listed in Table \ref{tab:elastic_constants}. These values satisfy the mechanical stability conditions for the tetragonal phase, confirming the structural integrity of ZnFe under elastic deformations. The combined evaluation of thermodynamic, dynamical, and mechanical stability provides strong theoretical evidence for the viability of ZnFe as a stable phase.

After establishing the FM ground state and confirming the structural stability of tetragonal ZnFe, we analyzed its electronic density of states (DOS), as shown in Fig. \ref{fig:6}(c) and \ref{fig:6}(d). The merging of valence and conduction bands in both spin channels indicates the metallic nature of ZnFe. The partial DOS (pDOS) further reveals that the Fe-$d$ and Zn-$d$ states predominantly contribute to the total DOS (TDOS). At the Fermi energy ($E_\mathrm{F}$), Fe-$d$ states dominate both the majority and minority spin channels, whereas the Zn-$d$ states make only minor contributions. The pronounced asymmetry between spin-up and spin-down states at $E_\mathrm{F}$ further substantiates the magnetic character of ZnFe. The computed total magnetization of ZnFe is 1.15 T, primarily originating from the Fe atoms, which contribute 2.43 $\mu_B$/atom.  

A key figure of merit for PMs is the maximum energy product, $(BH)_{\max}$, which quantifies the maximum magnetic energy density that a material can store per unit volume. As expressed in Eq. \ref{eqn:E3}, $(BH)_{\max}$ scales quadratically with magnetization, making it a crucial parameter for evaluating the performance of potential PM materials. Currently, the strongest commercially available PMs are RE-based intermetallic compounds such as Nd$_2$Fe$_{14}$B and SmCo$_5$, which can achieve maximum energy products of up to 445.7 kJ/m$^3$ in ideal conditions \cite{coey2010magnetism}. However, in practical applications, state-of-the-art PMs typically exhibit $(BH)_{\max}$ values around 350 kJ/m$^3$ \cite{westmoreland2020atomistic}. For ZnFe, we calculate a $(BH)_{\max}$ value of 264 kJ/m$^3$, which exceeds that of the $\tau$-MnAl alloy (101 kJ/m$^3$) \cite{islam2022origin}. The high energy product of ZnFe highlights its potential as a promising candidate for RE-free PMs, offering a competitive performance level in the absence of RE elements or costly metals such as Pt. 

A substantial magnetocrystalline anisotropy, characterized by a positive MAE with an out-of-plane magnetization orientation, is a critical requirement for PMs, as it determines the stability of magnetization and enhances magnetic coercivity, thereby increasing resistance to demagnetization. ZnFe exhibits uniaxial anisotropy, with an anisotropy energy density ($K$) of 0.76 MJ/m$^3$.  Another key parameter in evaluating PMs is the magnetic hardness parameter, $\kappa$, defined in Eq. \ref{eqn:E4}. This dimensionless quantity quantifies a material's potential to be developed into a compact PM given a sufficient saturation magnetization ($M_s$). The value of $\kappa$ is commonly used to classify ferromagnetic materials into hard, semi-hard, and soft magnets as discussed in section \ref{metho}. Our calculations show that ZnFe exhibits a $\kappa$ value of 0.85, placing it in the semi-hard category and highlighting its potential as a gap magnet, particularly in applications where RE-free alternatives are sought. The anisotropy field, defined as $H_a = \frac{2K}{M_s}$, establishes an upper bound on the achievable coercivity ($H_c$). However, it has been observed that the actual $H_c$ value is typically reduced to approximately one-fourth of $H_a$, a phenomenon known as the \enquote{Brown paradox} \cite{goll2000high}. In our study, ZnFe exhibits anisotropy fields exceeding 1 T, highlighting its potential for high coercivity.

The Curie temperature defines the critical threshold beyond which spontaneous bulk magnetization and long-range magnetic order vanish due to a magnetic order-disorder transition. For a PM to be practically viable, it must retain its magnetic properties under operational conditions, particularly at elevated temperatures. The computed T$_{\mathrm{C}}$ value for ZnFe (1230 K) satisfies the fundamental criterion for high-performance PMs, as outlined by Coey, which requires T$_{\mathrm{C}}$ to exceed 550 K \cite{COEY2012524}.  To gain deeper insight, we analyzed the exchange coupling constants $J_{ij}$. Positive values of $J_{ij}$ indicate FM interactions, whereas negative values signify AFM coupling. The Heisenberg exchange coupling parameters for ZnFe are depicted in Fig. \ref{fig:6}(e), computed for a cluster with a radius extending up to $7a$ (where $a$ is the optimized lattice parameter). Minimal changes were observed as the cluster radius increased beyond 2.5$a$ for ZnFe, indicating that the calculations have converged sufficiently. As expected, the strongest exchange interactions occur between the nearest-neighbor Fe atoms, exhibiting a maximum magnitude of approximately 30 meV and reinforcing FM coupling in ZnFe. In contrast, interactions involving non-magnetic Zn atoms, both Zn-Zn and Zn-Fe, are weak and nearly negligible across all distances. This confirms that the dominant magnetic interactions in ZnFe arise primarily from Fe-Fe coupling, which governs its high T$_{\mathrm{C}}$.

\subsection{Structural stability and magnetic properties of Fe$_8$N}

\begin{figure*}[t]
    \centering
    \includegraphics[width=1\linewidth]{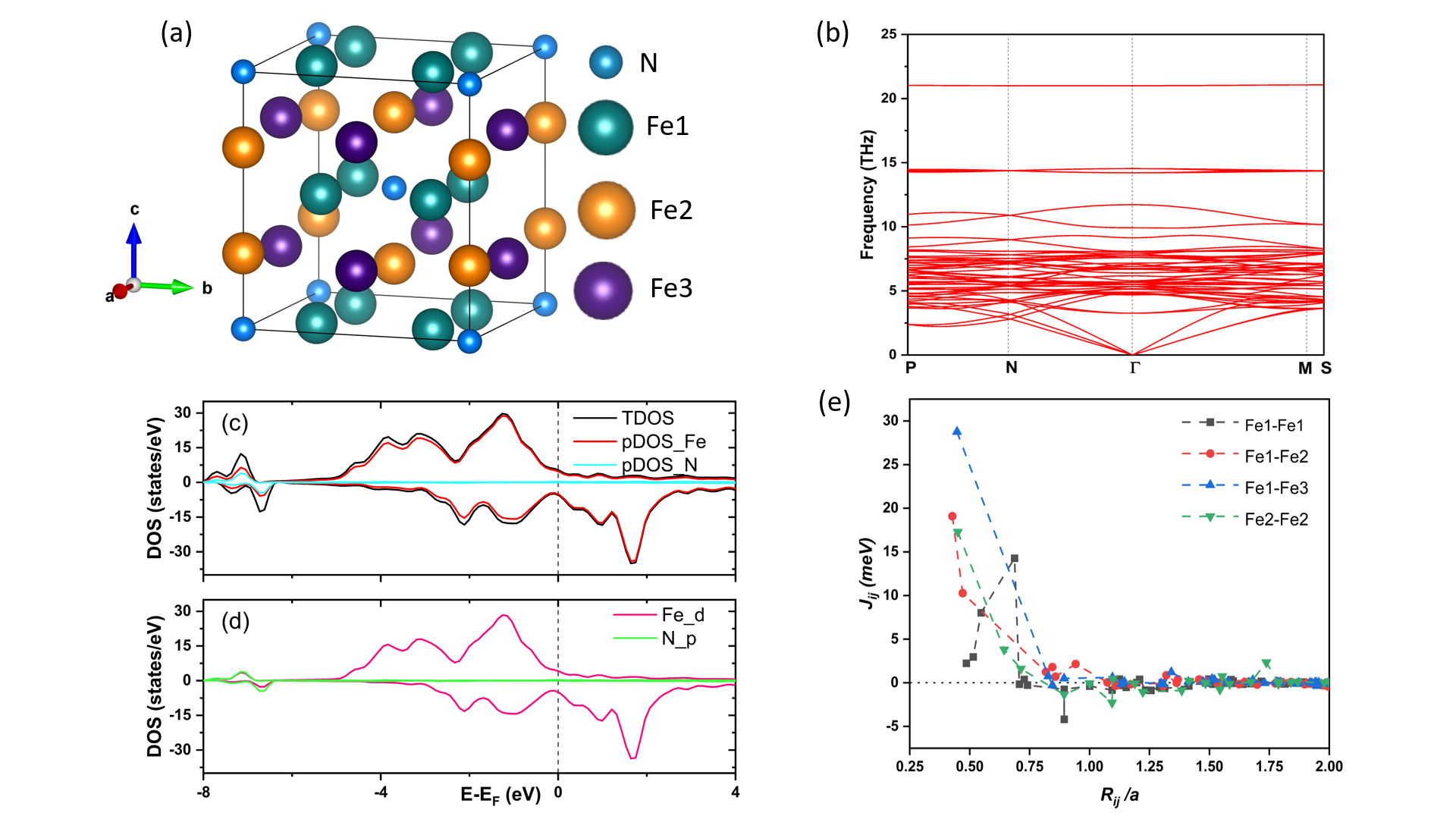}
    \caption{(a) Tetragonal unit cell of Fe$_8$N. (b) Phonon dispersion curves of Fe$_8$N, confirming its dynamical stability due to the absence of imaginary frequencies. (c) Spin-polarized total density of states (TDOS) and atom-projected partial density of states (pDOS) of Fe$_8$N, with the Fermi level ($E_F$) as the reference. (d) Orbital-resolved contributions of Fe $d$-orbitals and N $p$-orbitals to the TDOS, highlighting their role in the electronic structure. (e) Heisenberg exchange coupling parameters ($J_{ij}$) as a function of the interatomic distance normalized by the lattice parameter ($R_{ij}/a$), providing insights into the magnetic interactions in Fe$_8$N.}
    \label{fig:7}
\end{figure*}

 As compared to other iron compounds, the nitrides possess a greater degree of magnetization than the iron oxides and are more cost-effective than ferromagnetic alloys such as FePt. The ability to tune the magnetic properties of Fe-N compounds by adjusting the nitrogen concentration further enhances their appeal for technological applications. 
 
The tetragonal unit cell of Fe$_8$N comprises of two formula units, incorporating three distinct types of Fe atoms in 8, 4, and 4 atomic proportions, as depicted in Fig. \ref{fig:7}(a). Determining the correct magnetic ground state is crucial, as incorrect assumptions can compromise the reliability of predicted structural, electronic, and magnetic properties. To validate the default FM configuration reported for Fe$_8$N in the Materials Project database \cite{jain2013commentary}, we systematically investigated multiple magnetic configurations, as detailed in Section~S2 of the Supplementary Information. Total energy calculations revealed that the FM state possesses the lowest energy among all considered configurations, thereby confirming it as the true magnetic ground state.

The formation enthalpy of Fe$_8$N was computed, yielding a negative value of -0.255 eV/f.u., which indicates its thermodynamic stability. To further assess its dynamical stability, phonon dispersion calculations were performed along high-symmetry paths in the Brillouin zone. The phonon spectrum, as shown in Fig. \ref{fig:7}(b), exhibits no imaginary frequencies, confirming that Fe$_8$N remains dynamically stable under small perturbations. Since Fe$_8$N adopts a tetragonal crystal structure like ZnFe, the mechanical stability is evaluated using the Born-Huang criteria \cite{born1996dynamical,SINGH2021108068} as: $C_{11}-C_{12} > 0$,  $2C_{13}^{2}< C_{33} (C_{11}+C_{12})$, $C_{44} > 0$, and  $2C_{16}^{2}< C_{66} (C_{11}-C_{12})$. The calculated elastic constants for Fe$_8$N, presented in Table \ref{tab:elastic_constants}, satisfy all the necessary mechanical stability criteria for a tetragonal structure. This confirms that Fe$_8$N is mechanically stable and exhibits resistance to elastic deformation.

To further investigate the electronic properties of Fe$_8$N, we examined its electronic density of states (DOS), as depicted in Fig. \ref{fig:7}(c) and Fig. \ref{fig:7}(d). The asymmetry between spin-up and spin-down states at the Fermi energy ($E_\mathrm{F}$) confirms the magnetic nature of the compound. The DOS analysis indicates a substantial overlap between the valence and conduction bands in both spin channels, signifying its metallic character. At $E_\mathrm{F}$, Fe-$d$ states predominantly contribute to both the majority and minority spin channels, while N-$p$ states have a minimal contribution. The calculated total magnetization of Fe$_8$N is 1.21 T, primarily arising from Fe atoms, with the three distinct Fe sites exhibiting magnetic moments of 2.34, 2.14, and 2.82 $\mu_B$/atom, respectively. Fe$_8$N demonstrates a $(BH)_{\max}$ value of 264 kJ/m$^3$, which far exceeds that of conventional PMs, including hard ferrites and Alnico alloys, whose $(BH)_{\max}$ values typically range between 40 and 80 kJ/m$^3$ 
\cite{SKOMSKI20163}.

Fe$_8$N exhibits uniaxial magnetocrystalline anisotropy, a fundamental characteristic for PMs as it governs the stability of magnetization and enhances coercivity. The computed anisotropy constant ($K$) for Fe$_8$N is 0.57 MJ/m$^3$, indicating a preference for out-of-plane magnetization, which is essential for maintaining strong magnetic performance and resistance to demagnetization. The magnetic hardness parameter is another crucial factor that encapsulates both magnetic anisotropy and magnetization, with $\kappa > 0.1$ serving as a general criterion for identifying PMs. Fe$_8$N exhibits $\kappa$ value of 0.70, suggesting their potential as gap magnets. For reference, well-established PMs such as Nd$_2$Fe$_{14}$B, SmCo$_5$, Sm$_2$Fe$_{17}$N$_3$, and Sm$_2$Co$_{17}$ exhibit $\kappa$ values of 1.54, 4.4, 2.17, and 1.89, respectively, while AlNiCo and FeNi alloys typically have $\kappa$ values in the range of 0.3$-$0.5 \cite{OCHIRKHUYAG2024119755,SKOMSKI20163}. Although the magnetic hardness of ZnFe and Fe$_8$N falls slightly below the threshold for hard magnets ($\kappa \geq 1$), this parameter can be further optimized through alloying or other material engineering approaches.  The anisotropy field, given by $H_a = \frac{2K}{M_s}$,  is essential for ensuring strong resistance to demagnetization, which is a key requirement for PM applications. In the case of Fe$_8$N, the calculated anisotropy field exceeds 1 T, indicating that Fe$_8$N could be a promising candidate for PM applications.  

For a material to function effectively as a PM, it must retain its magnetic properties well above room temperature. Our computed T$_{\mathrm{C}}$ for Fe$_8$N is 1585 K, which far exceeds the minimum threshold of 550 K established by Coey for high-performance PMs \cite{COEY2012524}. To gain deeper insight into the magnetic interactions governing T$_{\mathrm{C}}$, we analyze the exchange coupling constants, $J_{ij}$, which provide a quantitative measure of the strength and nature of magnetic interactions between atomic pairs. Fig. \ref{fig:7}(e) presents the Heisenberg exchange coupling parameters for Fe$_8$N. Given that Fe$_8$N comprises three crystallographically distinct Fe sites (Fe1, Fe2, and Fe3), numerous Fe–Fe exchange interactions are present; however, those with significantly weaker coupling strengths have been omitted from Fig.~\ref{fig:7}(e) for clarity. Among the interactions, the Fe1–Fe3 coupling is the most dominant at shorter interatomic distances, exhibiting a strength of approximately 29~meV. This is followed by notable Fe1–Fe2 and Fe2–Fe2 interactions, which also play a significant role in stabilizing the ferromagnetic order.

\section{Discussions}

The search for RE-free PMs has gained momentum due to the growing demand for cost-effective and sustainable alternatives. Although conventional ferrite and AlNiCo magnets offer moderate performance, they fall significantly short of those of Nd-Fe-B and Sm-Co in terms of energy density and magnetic hardness. In this study, we employ a data-driven strategy integrated with first-principles computational screening to discover new materials that bridge this performance gap, aiming to achieve optimal cost-to-performance efficiency. Our multi-step screening process identified several RE-free binary compounds with high saturation magnetization, magnetocrystalline anisotropy, and Curie temperature. Additionally, a thorough review of existing literature was conducted to filter out already known compounds for PMs and to distinguish them from lesser-known or unexplored compounds without reported synthesis or stability challenges. Through this process ZnFe and Fe$_8$N are identified as strong candidates for PM applications.  It is important to note that Fe (present in both these compounds) is the most critical ferromagnetic element, valued not only for its high magnetic moment and Curie temperature but also for its natural abundance, a reason for its pivotal role in magnetic industry.

Identifying the magnetic ground state is the most urgent problem to be solved for predictive high-throughput screening on magnetic materials, since the magneto-structural coupling can significantly alter the energy landscape. We rechecked the magnetic state of the above mentioned compounds by performing explicit calculations with different magnetic ordering. However, the number of possible AFM configurations which should be considered to define the magnetic ground state can be practically indefinitely large, depending on the crystal structures and the magnetic ions involved and cell size employed for calculation. For the tetragonal unit cell chosen for ZnFe (Fig. \ref{fig:6}(a)) and Fe$_8$N (Fig. \ref{fig:7}(a)), the FM ground state is found to be the lowest energy state, which also exhibited a negative enthalpy of formation, along with dynamical and mechanical stability. Notably, all known iron nitrides, whether in bulk form, as thin films, or as nanoparticles, are inherently metastable due to kinetic constraints.  The ordered phases of these materials exist only within a moderate temperature window of 350–550 $^{\circ}$C. Due to these inherent limitations, iron nitrides have been relatively less explored compared to their iron oxide counterparts, despite exhibiting superior magnetic properties \cite{bhattacharyya2015iron}. In our study, ZnFe and Fe$_8$N exhibit energies above the convex hull of 23 meV/atom and 10 meV/atom, respectively. As both values lie well within the commonly accepted metastability threshold of 50 meV/atom, these compounds are considered promising candidates for experimental synthesis. Moreover, the chemical similarity of isoelectronic elements suggests that partial or full elemental substitution within these compounds could stabilize the phase while potentially enhancing magnetic properties. Further, the stability and performance of metastable materials can often be improved by modifying the experimental processes. For instance, the use of reactive precursors such as NH$_3$ can lower the thermodynamic barrier, enabling the formation of previously inaccessible metastable phases \cite{sun2017thermodynamic}.  Furthermore, non-equilibrium methods such as molecular beam epitaxy, melt spinning, mechanical alloying, and  specialized nanostructuring techniques can be employed to achieve metastable phases, as demonstrated in the $\varepsilon$ phase of MnAl \cite{jimenez2012exchange}.  Despite ongoing discussions regarding the precise saturation magnetization of $\alpha''$-Fe$_{16}$N$_2$, as well as challenges in optimizing its bulk magnetic properties, this compound remains a highly attractive candidate for PM applications due to its cost-effectiveness and the abundance of raw materials. 

The saturation magnetization values of ZnFe and Fe$_8$N surpass that of conventional hard ferrites ($\sim$0.40 T) \cite{MOHAPATRA20181, SKOMSKI20163} and also exceed those of widely studied gap magnets such as MnAl (0.82 T) \cite{pareti1986magnetic, wei2014tau} and MnBi (0.80$-$0.93 T) \cite{PhysRevB.83.024415, YANG2021157312, guo1992magnetic}. While these results are promising, further investigations into alloying effects may lead to additional improvements in saturation magnetization, though such an analysis is beyond the scope of this study. Although the values of magnetic hardness parameter are slightly below the threshold for hard magnets (\(\kappa \geq 1\)), targeted alloying strategies could potentially optimize their magnetic anisotropy and coercivity, thereby enhancing their overall performance. The high Curie temperatures of ZnFe (1230 K) and Fe$_8$N (1585 K) further reinforce their potential for high-temperature applications. Despite these advantages, a key limitation of iron nitrides is their relatively low decomposition temperature, which restricts traditional microstructural enhancements. A potential strategy to address this challenge involves the large-scale synthesis of high-purity $\alpha''$-Fe$_{16}$N$_2$ powder, which could facilitate the fabrication of bonded magnets. Nevertheless, further research is required to enhance the coercivity and thermal stability of these materials to meet the demands of practical applications. Ultimately, our study highlights the effectiveness of high-throughput computational screening in the discovery of new PM materials. This study prioritizes the selection of elements based on their intrinsic magnetic properties, economic viability, resource abundance, and environmental sustainability.

\section{Conclusions}

  In this study, we employed a  materials-specific screening method integrated with a  high-throughput data-driven approach to search for potential candidates within the known crystal structures from the Materials Project database. The initial screening focused on binary compounds containing at least one 3$d$ transition metal paired with a 3$d$, 4$d$, 5$d$, or $p$-block element (excluding halogens, noble gases, rare-earth elements, and oxides) in hexagonal or tetragonal crystal structures. Magnetization and thermodynamic stability data guided the initial filter, while compounds with costly elements like Pt and Au were excluded. Shortlisted candidates were further subjected to detailed first-principles calculations to evaluate their magnetocrystalline anisotropy energy  and Curie temperature. Stability was assessed using conventional metrics, including formation energy, hull distance, phonon band dispersion, and mechanical stability criteria. Among the promising candidates, ZnFe and Fe$_8$N were screened with hull distances well within the metastability threshold of 50 meV/atom. Both exhibit uniaxial magnetic anisotropy (0.76 MJ/m$^3$ for ZnFe and 0.57 MJ/m$^3$ for Fe$_8$N), high saturation magnetization values (1.15 T and 1.21 T, respectively), and notable magnetic hardness parameters ($\kappa = 0.85$ and $0.70$), classifying them as potential gap magnets. Their magnetic order is driven by strong Fe–Fe exchange couplings, resulting in high T$_{\mathrm{C}}$ values of 1230 K (ZnFe) and 1585 K (Fe$_8$N), indicating suitability for high-temperature applications. It is worth noting that ZnFe has thus far been investigated only through this particular study, necessitating further experimental validation. However, first principles based analysis, such as this, provide valuable guidance for experimentalists, aiding in the identification of promising candidates and offering insights into their structural and magnetic properties. Although the synthesis of these compounds remains an open question, the answer to that depends, among other factors, on the kinetics of the synthesis route.  The materials identified in this offer a promising foundation for the design and development of next-generation rare-earth-free permanent magnets, with potential for further optimization through compositional tuning and tailored synthesis routes.
 
\section{Acknowledgements}

The high performance computational facilities viz. Aron (AbCMS lab, IITB), Dendrite (MEMS dept., IITB), Spacetime, IITB and CDAC Pune (Param Yuva-II) are acknowledged for providing the computational hours. JJ acknowledges funding received through PMRF grant (PMRF ID : 1300138). AB thanks funding received through BRNS regular grant (BRNS/37098) and SERB power grant (SPG/2021/003874).

\bibliography{prop}

\bibliographystyle{apsrev4-2} 

\end{document}